\pdfoutput=1
\RequirePackage{ifpdf}
\ifpdf 
\documentclass[pdftex]{sigma}
\else
\documentclass{sigma}
\fi

\begin{document}

\allowdisplaybreaks

\renewcommand{\thefootnote}{$\star$}

\renewcommand{\PaperNumber}{071}

\FirstPageHeading

\ShortArticleName{Conservation Laws, Hodograph Transformation and Boundary Value Problems}

\ArticleName{Conservation Laws, Hodograph Transformation\\ and Boundary Value Problems of Plane Plasticity\footnote{This
paper is a contribution to the Special Issue ``Geometrical Methods in Mathematical Physics''. The full collection is available at \href{http://www.emis.de/journals/SIGMA/GMMP2012.html}{http://www.emis.de/journals/SIGMA/GMMP2012.html}}}

\Author{Sergey I.~SENASHOV~$^\dag$ and Alexander YAKHNO~$^\ddag$}

\AuthorNameForHeading{S.I.~Senashov and A.~Yakhno}

\Address{$^\dag$~Siberian State Aerospace University, Krasnoyarsk, Russia}
\EmailD{\href{mailto:sen@sibsau.ru}{sen@sibsau.ru}}

\Address{$^\ddag$~Departamento de Matem\'{a}ticas, CUCEI, Universidad de Guadalajara, 44430, Mexico}
\EmailD{\href{mailto:alexander.yakhno@cucei.udg.mx}{alexander.yakhno@cucei.udg.mx}}

\ArticleDates{Received April 18, 2012, in f\/inal form September 29, 2012; Published online October 13, 2012}

\Abstract{For the hyperbolic system of quasilinear f\/irst-order partial dif\/ferential equations, linearizable by hodograph transformation, the conservation laws are used to solve the Cauchy problem. The equivalence of the initial problem for quasilinear system and the problem for conservation laws system permits to construct the characteristic lines in domains, where Jacobian of hodograph transformations is equal to zero. Moreover, the conservation laws give all solutions of the linearized system. Some examples from the gas dynamics and theory of plasticity are considered.}


\Classification{35L65; 58J45; 74G10}

\renewcommand{\thefootnote}{\arabic{footnote}}
\setcounter{footnote}{0}

\section{Introduction}
\label{sec:Introduction}

The Riemann method is one of the classical ones and is used widely for solution of Cauchy problem for hyperbolic linear equation of second order for one function of two independent variables
\begin{gather}
\label{0}
 \frac{\partial^2 u}{\partial x\partial y} + a(x,y)\frac{\partial u}{\partial x} + b(x,y)\frac{\partial u}{\partial y} + c(x,y) u = f(x,y),
\end{gather}
where $u(x,y)$ is specif\/ied on the boundary $L = \{(x,y)\,|\, x=x(\tau), y=y(\tau)\}$, which is not a~cha\-racteristic line.

To determine the value of $u$ at point $x=x'$, $y=y'$, the so-called Riemann function $R(x,y;x',y')$ is used, which is a solution of the special characteristic Cauchy problem (Goursat problem) for the homogeneous adjoint equation, corresponding to~(\ref{0})
\begin{gather}
\label{adjoint}
 \frac{\partial^2 R}{\partial x\partial y} -\frac{\partial (a R)}{\partial x} - \frac{\partial (b R)}{\partial y} + cR = 0, \\
 \left.\left(\frac{\partial R}{\partial y} - aR \right)\right|_{x=x'}=0, \qquad \left.\left(\frac{\partial R}{\partial x} - bR \right)\right|_{y=y'}=0, \qquad \left.R \right|_{x=x',\, y=y'}=1. \nonumber
\end{gather}
If the Riemann function is determined, then the solution of Cauchy problem for~(\ref{0}) is given by an integral representation. The main reason for the introduction of adjoint equation~(\ref{adjoint}) is to make the line integral vanish around closed paths~\cite{Martin:1951}.
In other words there is a conservation law of the special form, that, for example in~\cite{Chirkunov:2009} is called the obvious one.

The same method is applied for linear system obtained from quasilinear one by hodograph transformation. This `speedgraph'  transformation, see for example~\cite{vonMises:1958}, is just an interchange of roles of the unknown functions and the independent variables. There is a lot of works, where the hodograph-type transformations are used to linearize dif\/ferent kinds of systems of dif\/ferent orders. Let us mention paper~\cite{Clarkson:1989}, where an algorithm for establishing whether a given quasilinear equation is linearizable (is solvable in terms of either a linear partial dif\/ferential equation or of a linear integral equation) is proposed to apply Painlev\'{e} test to quasilinear equations. The so-called extended hodograph transformation (including integral term) is introduced to reduce quasilinear PDE to the semi-linear one.
In work~\cite{Fushchych:1993} the hodograph transformation with the principle of nonlinear superposition is used to produce some new solutions of the special classes of equations.

But the linearization of quasilinear systems considered later by hodograph transformation is valid only in the domains, where the Jacobian of transformation is not equal to zero, which is unknown before the solution of the quasilinear system, because of the dependence of characteristic curves on the solution.

Conservation laws of quasilinear homogeneous hyperbolic system for two functions of two independent variables, related to the solution of corresponding linearized system are used in the paper for the solution of the Cauchy problem. Vanishing of Jacobian is not a restriction now, this allows to construct the characteristic f\/ields corresponding to the simple waves.

The paper is structured as follows. In Section~\ref{sec:system} we stand the problem and describe some basic properties of a hyperbolic quasilinear system. Section~\ref{sec:CL} deals with the construction of the conservation laws of the special form, its relation with a solution of the linearized system and the description of the solution of Cauchy problem. Some applications of the exposed method are considered in Section~\ref{sec:Ex}, in particular the Cauchy problem from the theory of plane plasticity for the loaded cavity is solved for any convex form of contour.

\section{Hyperbolic quasilinear system}\label{sec:system}

Let us consider a quasilinear system of homogeneous PDEs of two independent variables $x$, $y$ and two dependent ones $u$, $v$ in the form~\cite{Rozhdestvenskii:1983}
\begin{gather}
	A \frac{\partial U}{\partial x} + B \frac{\partial U}{\partial y} = 0,
	\label{eq0}
\end{gather}
 where $A = \left\|a_{ij}(u,v)\right\|$, $B = \left\|b_{ij}(u,v)\right\|$, $i,j = 1,2$, $U = (u,v)^{T}$.

If matrix $A$ is not degenerate, then system (\ref{eq0}) can be written in the normal form
\begin{gather}
	 \frac{\partial U}{\partial x} + M \frac{\partial U}{\partial y} = 0,
	\label{eq00}
\end{gather}
 where $M=\left\|m_{ij}(u,v)\right\|$.

Let us set up the Cauchy problem for system~(\ref{eq00}) due to~\cite{Rozhdestvenskii:1983}: in some neighborhood of the arc $C$: $a \leqslant \tau \leqslant b$ of an initial curve $L = \{(x,y):x=x(\tau), y=y(\tau)\}$ it is necessary to determine the solution of~(\ref{eq00}) satisfying the initial condition on~$L$
\begin{gather}
\label{init_conds_nonl}
U(x(\tau),y(\tau)) = U^0(\tau),\qquad \tau\in[a,b].
\end{gather}
The Cauchy problem~(\ref{eq00}),~(\ref{init_conds_nonl}) is supposed to be the normal one (in the sense of~\cite{Rozhdestvenskii:1983}) and the solution exists in some neighborhood of~$L$.

Let system (\ref{eq00}) be a strictly hyperbolic one. It means that matrix~$M$ has two real dif\/ferent eigenvalues $\lambda_1$ and $\lambda_2$, obtained as  roots of the equation{\samepage
\[
\det(M - \lambda E) = 0 \quad \Rightarrow \quad 2\lambda_{1,2} = m_{11} + m_{22} \pm \sqrt{(m_{11} - m_{22})^2 + 4 m_{12}m_{21}},
\]
that gives two left eigenvectors $l_1 = \left(l_1^1,l_1^2\right)$ and $l_2 = \left(l_2^1,l_2^2\right)$ respectively.}

Let us consider the dif\/ferential forms
\[
\omega_k = l_k^1(u,v) du + l_k^2(u,v) dv = 0, \qquad k=1,2,
\]
which can be integrated, because in this case there always exist integrating factors. Then, the corresponding two integrals $\Phi_k(u,v) = \text{const}$ can be taken as Riemann invariants $r_k = \Phi_k(u,v)$ and system~(\ref{eq00}) takes a diagonal form
\begin{gather}
 \frac{\partial R}{\partial x} + \Lambda \frac{\partial R}{\partial y} = 0, \qquad \Lambda = \operatorname{diag}(\lambda_1,\lambda_2), \qquad  R = (r_1,r_2)^{T}.
\label{Riemann}
\end{gather}
System (\ref{Riemann}) has two families of real characteristic curves determined by the following equations
\begin{gather}
	\frac{\mathrm{d} y}{\mathrm{d} x} = \lambda_1(u, v), \qquad \frac{\mathrm{d} y}{\mathrm{d} x} = \lambda_2(u, v), \qquad \lambda_1 \ne \lambda_2,
\label{char-curves}
\end{gather}
that is, variable~$r_1$ is the invariant along the f\/irst characteristic curve and~$r_2$ is the invariant along the second one. For more detailed theory of Riemann invariants see, for example,~\cite{Jeffrey:1976,Peradzynski:1985,Rozhdestvenskii:1983}.

The systems of such a form are widely used in the mechanics of a continuum media~\cite{Rozhdestvenskii:1983}, in the gas dynamics for describing isoentropic plane-symmetry f\/lows; in the theory of plane plasticity for the stresses of a deformed region under the dif\/ferent yield criterions~\cite{Kachanov:2004}, for the motion of granular materials, for the propagation of the plane wave of loading in homogeneous semi-inf\/inite elastic-plastic beam~\cite{Nowacki:1978}, etc.

It is well known, that system (\ref{eq00}) can be linearized by a so-called \textit{hodograph transformation} of the form $x = x(u, v)$, $y = y(u, v)$. Thus, system (\ref{eq0}) takes a linear form
\begin{gather*}
\det(A) \big(A^T\big)^{-1} \nabla_U y(u,v) = \det(B) \big(B^T\big)^{-1} \nabla_U x(u,v), \qquad  \nabla_U = \left({\partial }/{\partial u}, {\partial }/{\partial v} \right)^T,
\end{gather*}
and for diagonal form (\ref{Riemann}) one can obtain
\begin{gather}
\label{hodograph}
\Lambda \nabla_R y(r_1,r_2) = \det(\Lambda) \nabla_R x(r_1,r_2), \qquad  \nabla_R = \left({\partial }/{\partial r_1}, {\partial }/{\partial r_2} \right)^T.
\end{gather}

Let us note, that it is possible to obtain the solution of system~(\ref{Riemann}) from the solution of~(\ref{hodograph}) and vice versa only when two corresponding Jacobians $J_1 = \left|{\partial(r_1,r_2)}/{\partial(x,y)}\right|$  and $J_2 = \left| {\partial(x,y)}/{\partial(r_1,r_2)} \right|$ are not equal to zero
\begin{gather*}
J_1 =  \frac{\partial r_1}{\partial y} \frac{\partial r_2}{\partial x} - \frac{\partial r_2}{\partial y} \frac{\partial r_1}{\partial x} = J^{-1}_2.
\end{gather*}

Let us mention, that all solutions of system~(\ref{Riemann}) can be divided into two classes: the singular and non-singular ones. The singular solutions are obtained as solutions of equations $J_i = 0$. And non-singular ones are the solutions of system~(\ref{hodograph}). Solutions with degenerate Jacobi matrix form a class of solutions called multiple waves and can be found by so-called method of the degenerate hodograph (see, for example~\cite{Meleshko:2005}), and in general belong to the class of partially-invariant solutions~\cite{Ovsyannikov}.  A solution of considered system (\ref{Riemann}) for which the rank of the Jacobi matrix in a domain $G\in \mathbb{R}^2$ satisf\/ies the condition
\[
\operatorname{rank} \frac{\partial (r_1, r_2)}{\partial(x,y)} = r
\]
is called a multiple wave of the rank $r$. In general $r < \min\left\{n,m\right\}$, where $n$ is a number of independent variables and $m$ is a number of functions. There are only two possibilities for the rank in our case: $r=0$ o $r=1$ ($r=2$ corresponds to non-singular solution). The case $r=0$, means that both Riemann invariants are constant functions, that corresponds to the constant functions $u$ and $v$. If $r=1$, then a multiple wave is called a simple wave and is widely known. For this kind of solution there are f\/inite relations between the functions $u$, $v$. For example, simple waves for the plane plasticity equations, considered later, were analyzed in~\cite{Khristianovich:1936}.

\begin{figure}[t]
\centering
\includegraphics[scale=0.2]{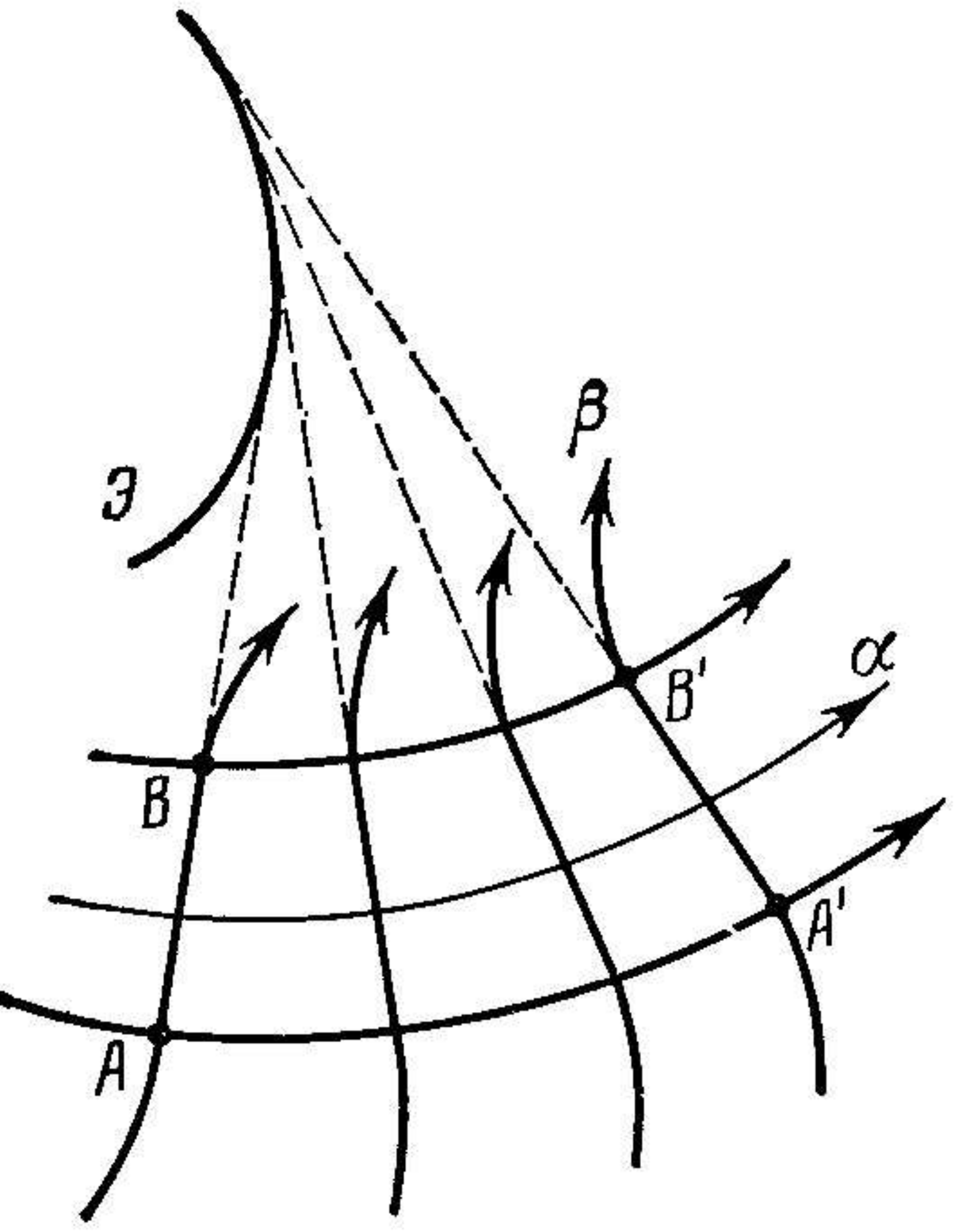}\quad \includegraphics[scale=0.2]{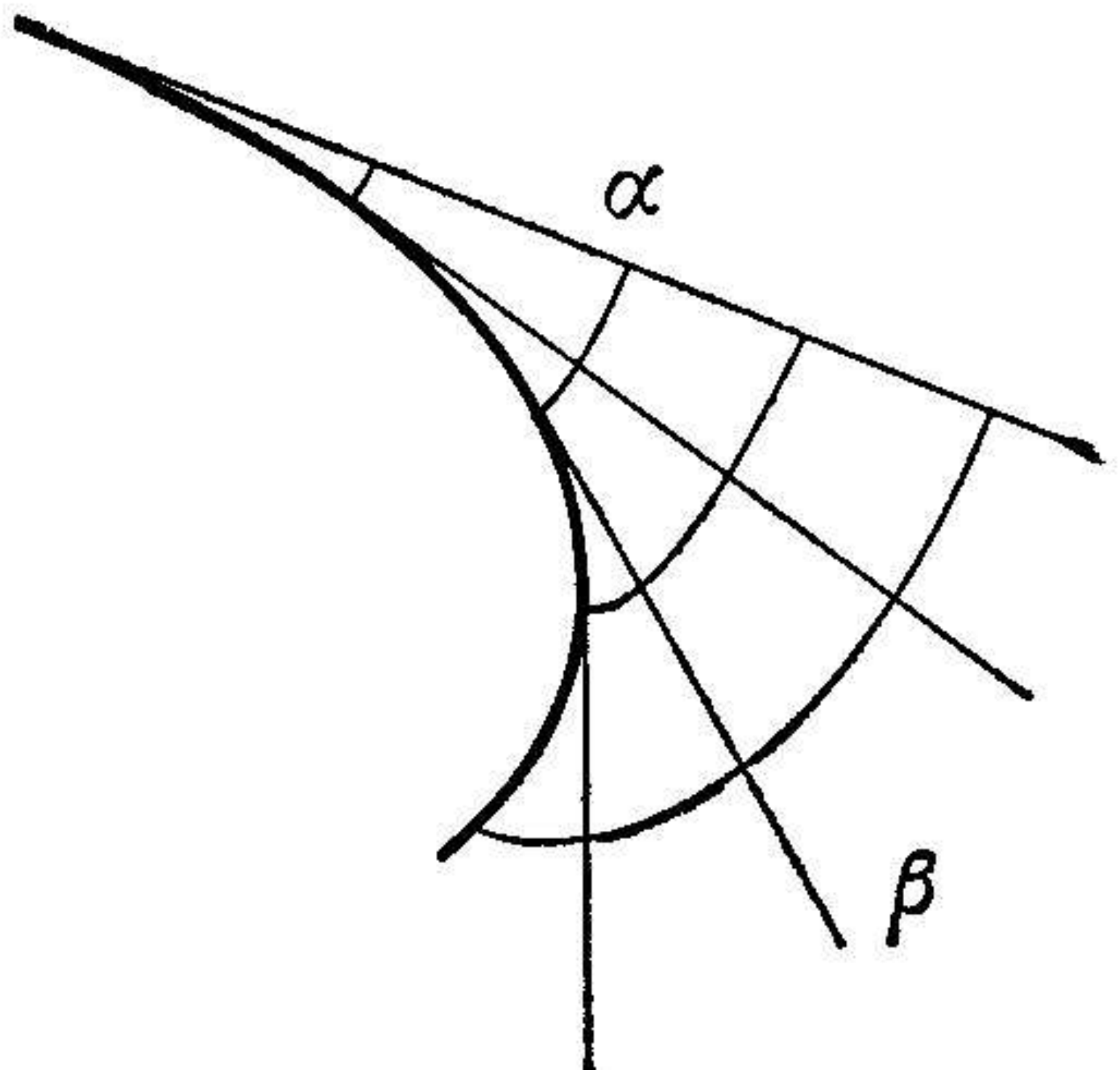}
\caption{Areas of straight line characteristics.}\label{f1}
\end{figure}

The main dif\/f\/icult in practical applications is the determination of a domain $G$, mentioned above, because the domain of determinacy of the Cauchy problem for the system of quasilinear equations is determined simultaneously with the solution and, generally speaking, cannot be indicated beforehand~\cite{Rozhdestvenskii:1983}. In other words, before the complete solution of Cauchy problem~(\ref{eq00}),~(\ref{init_conds_nonl}) it is impossible to calculate both~$J_i$. By the same reason,  before the complete solution it is dif\/f\/icult to point out the domain, where the characteristic curves (\ref{char-curves}) do convert to straight lines (see Fig.~\ref{f1}) for given form of the initial curve and for given initial conditions.

Due to the linearity of~(\ref{hodograph}) any non-singular solutions can be obtained from the one by the superposition principle (that corresponds to the symmetry, admitted by a linear system). In the theory of group analysis such kind of systems is called automorphic: if one particular solution is known, then the orbit of this solution under the group of admitted symmetries forms locally a general solution of the given system. In such a way, for example, some new exact solutions for the plane plasticity system were constructed in~\cite{Yakhno:2010}. Using conservation laws permits to f\/ind out both nonsingular and singular solutions.

System~(\ref{eq00}) can be extended for three or more functions and in this case is called one-dimensional system of hydrodynamic type~\cite{Tsarev}. A natural hamiltonian formalism was proposed for this class of homogeneous systems of PDE  and the generalized hodograph method  generates from every its solution a symmetry (commuting f\/low) that f\/inally leads to solution. Another approach to generalize the concept of Riemann invariants to systems with more than two independent variables ($m>2$) or with the number of functions more than two ($n>2$), is related with so-called simple integral elements~\cite{Grundland:1984}.

\section{Conservation laws}\label{sec:CL}

The concept of conservation laws is one of the fundamental characteristics of a real physical process.  One can f\/ind the formal def\/inition of the conservation laws of the system of dif\/fe\-ren\-tial equations, for example in~\cite{Vinogradov:1999}, where the concept of the so-called operator of universal linearization is used.

\begin{figure}[t]
\centering
\includegraphics[scale=0.45]{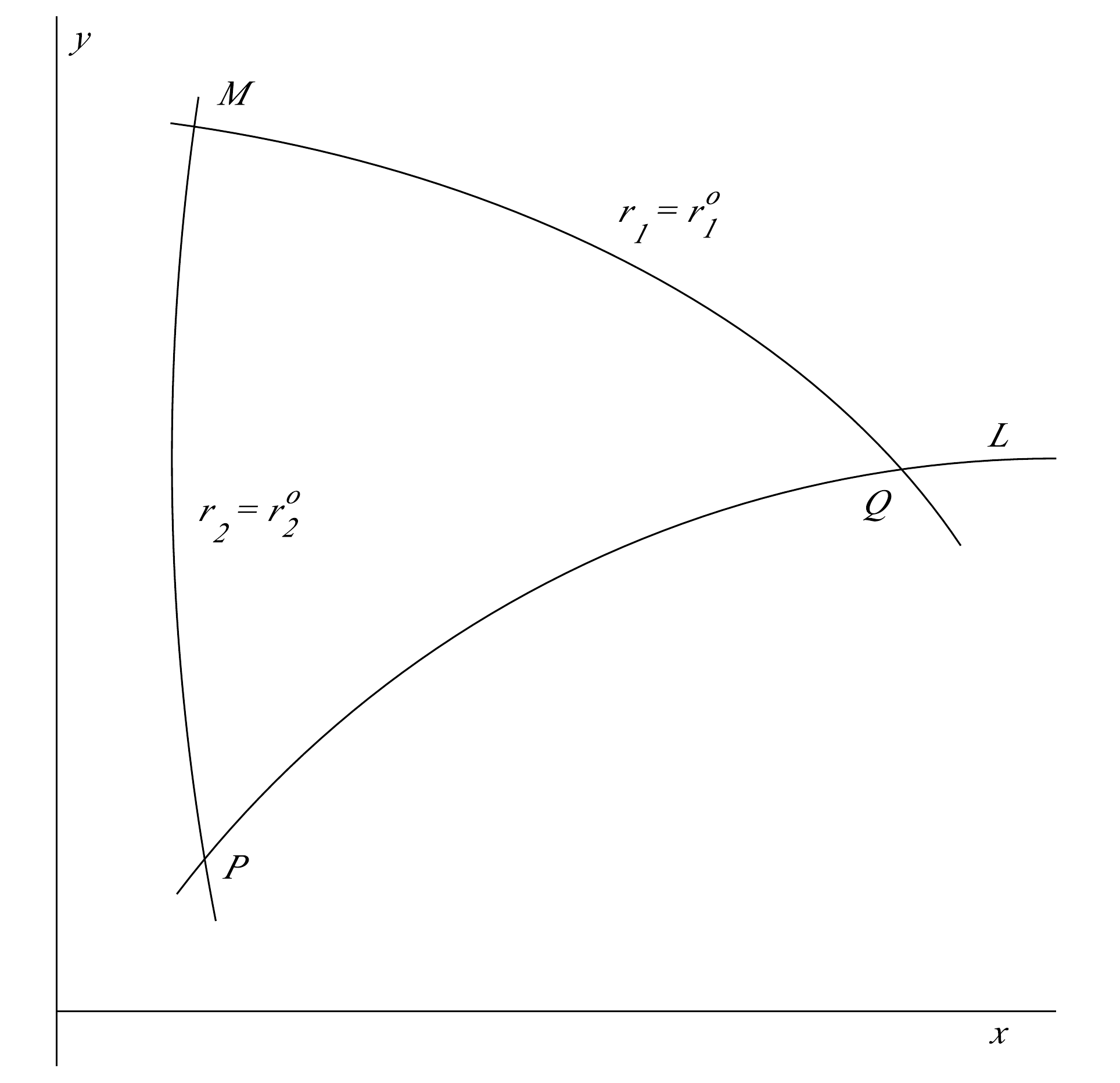}

\caption{Cauchy problem.}
\label{f2}
\end{figure}

Let us seek the conservation law of system (\ref{Riemann}) directly in the form
\begin{gather}
\label{CL1}
 \frac{\partial}{\partial x}  \varphi(u,v) + \frac{\partial }{\partial y} \psi(u,v)= 0,
\end{gather}
which is valid for any solution of (\ref{Riemann}). Let us multiply (\ref{Riemann}) by a vector $\alpha = (\alpha_1,\alpha_2)$ \cite{Rozhdestvenskii:1983}, then, eliminating $\alpha$, we obtain the linear system for functions $\varphi$ and $\psi$
\begin{gather}
\label{CL2}
\Lambda \nabla_R \varphi(r_1,r_2) = \nabla_R \psi(r_1,r_2).
\end{gather}
Now let us describe the way to solve Cauchy problem using conservation laws. Let $P(x(a),y(a))$, $Q(x(b),y(b))$ be two end-points of the arc $C$, $M(M_x, M_y)$ be a point of intersection of two characteristic lines: $r_1 = r_1^0$, going from the point~$Q$ and $r_2 = r_2^0$, going from the point~$P$ (see Fig.~\ref{f2}).

Let us apply the Riemann method not for linearized system~(\ref{hodograph}), but for system~(\ref{CL2}). Taking integral over the closed path~$PQM$, which due to Green theorem is equal to zero, we have
\begin{gather*}
\oint_{PQM}{\psi dx - \varphi dy} = 0 = \int_{PQ} {\psi dx - \varphi dy}  + \int_{r_1 = r_1^0}\left(\psi  - \varphi \lambda_1 \right)dx +  \int_{r_2 = r_2^0}\left(\psi  - \varphi \lambda_2 \right) dx.
\end{gather*}
Integrating by parts the above two integrals
\begin{gather*}
\int_{r_1 = r_1^0}\left(\psi  - \varphi \lambda_1 \right)dx = \int_{QM}\left(\psi  - \varphi \lambda_1 \right)dx =  x \left(\psi  - \varphi \lambda_1 \right)\big|_{x=Q_x}^{x=M_x} - \int_{QM} x d\left(\psi  - \varphi \lambda_1 \right),
\\
\int_{r_2 = r_2^0}\left(\psi  - \varphi \lambda_2 \right)dx = \int_{MP}\left(\psi  - \varphi \lambda_2 \right)dx =   x \left(\psi  - \varphi \lambda_2 \right)\big|_{x=M_x}^{x=P_x} - \int_{MP} x d\left(\psi  - \varphi \lambda_2 \right),
\end{gather*}
and taking (without loss of generality) the following conditions
\begin{gather}
\label{init_conds_linear}
\left. \left(\psi - \lambda_1 \varphi \right)\right|_{r_1 = r_1^0} = 1, \qquad \left. \left(\psi - \lambda_2 \varphi \right)\right|_{r_2 = r_2^0} = 0,
\end{gather}
we get coordinate $M_x$
\begin{gather}
\label{sol_x}
M_x = x(b) -  \int_{PQ} \left(\psi \frac{dx}{d\tau} - \varphi \frac{dy}{d\tau}\right)d\tau.
\end{gather}

Analogically, for the $y$-coordinate of point $M$ we obtain
\begin{gather}
\label{sol_y}
M_y = y(b) -  \int_{PQ} \left(\psi \frac{dx}{d\tau} - \varphi \frac{dy}{d\tau}\right)d\tau,
\end{gather}
but now the conditions for the functions $\varphi$, $\psi$ are the following ones
\begin{gather}
\label{init_conds_linear_y}
\left. \left(\psi/ \lambda_1 - \varphi \right)\right|_{r_1 = r_1^0} = 1, \qquad \left. \left(\psi/\lambda_2 - \varphi \right)\right|_{r_2 = r_2^0} = 0.
\end{gather}
In such a way, the coordinates of point $M$ are known and one can reconstruct the values of functions $u$, $v$ from the initial condition~(\ref{init_conds_nonl}).

Finally, we have demonstrated the following
\begin{theorem}
\label{th:1}
The solution of Cauchy problem for quasilinear system \eqref{eq00}, \eqref{init_conds_nonl} is equivalent to the solution of two Cauchy problems for the linear system \eqref{CL2} with conditions \eqref{init_conds_linear} and  \eqref{init_conds_linear_y}.
\end{theorem}

The main application of Theorem \ref{th:1} is that one can obtain the solution of Cauchy problem for the original system without any restrictions on the Jacobian of hodograph transformation.

Note, that the above procedure can be used to solve the characteristic problem, when the initial data are prescribed on two intersecting characteristics (see~\cite{Senashov:2004}).

Let us note, that in~\cite{Chirkunov:2009} the classif\/ication of second-order linear dif\/ferential equations with two independent variables in terms of f\/irst-order conservation laws is proposed. System~(\ref{Riemann}) can be reduced to such an equation, and in this case conservation laws~(\ref{CL1}) are among the f\/irst-order ones.

Let us express (\ref{hodograph}) and (\ref{CL2}) in terms of the hyperbolic equations of the second order
\begin{gather}
\label{x}
\frac{\partial^2 x}{\partial r_1 \partial r_2} - \frac{1}{\lambda_1 - \lambda_2} \frac{\partial \lambda_2}{\partial r_2} \frac{\partial x}{\partial r_1} + \frac{1}{\lambda_1 - \lambda_2} \frac{\partial \lambda_1}{\partial r_1} \frac{\partial x}{\partial r_2} = 0,
\\
\label{phi}
\frac{\partial^2 \varphi}{\partial r_1 \partial r_2} + \frac{1}{\lambda_1 - \lambda_2} \frac{\partial \lambda_1}{\partial r_2} \frac{\partial \varphi}{\partial r_1} - \frac{1}{\lambda_1 - \lambda_2} \frac{\partial \lambda_2}{\partial r_1} \frac{\partial \varphi}{\partial r_2} = 0.
\end{gather}

The most general equivalence transformations, preserving the dif\/ferential structure of these equations have the form~\cite{Ovsyannikov}
\begin{gather}
\label{R}
R_i = f_i(r_i),  \qquad u(r_1,r_2) = w(r_1,r_2) V(R_1,R_2), \qquad i=1,2.
\end{gather}
Using this concept, we can demonstrate the following
\begin{proposition}
The conservation laws~\eqref{CL2} give all nonsingular solutions of linearized system~\eqref{hodograph}.
\end{proposition}

\begin{proof}
Let us consider the well-known lemma about Laplace invariants $\left(h^{(i)},k^{(i)}\right)$ for two equivalent hyperbolic equations \cite{Ovsyannikov}, namely, two hyperbolic equations of the form
\begin{gather*}
\frac{\partial^2 V^{(1)}}{\partial r_1 \partial r_2} + C^{(1)}_1(r_1,r_2)\frac{\partial V^{(1)}}{\partial r_1} + C^{(1)}_2(r_1,r_2)\frac{\partial V^{(1)}}{\partial r_2} + C^{(1)}_3(r_1,r_2) V^{(1)}=0, \\
\frac{\partial^2 V^{(2)}}{\partial R_1 \partial R_2} + C^{(2)}_1(R_1,R_2)\frac{\partial V^{(2)}}{\partial R_1} + C^{(2)}_2(R_1,R_2)\frac{\partial V^{(2)}}{\partial R_2} + C^{(2)}_3(R_1,R_2) V^{(2)}=0
\end{gather*}
are equivalent if\/f there exist a functions $f_i$, $w(r_1,r_2)$ (\ref{R}), such that its corresponding Laplace invariants
\begin{gather*}
h^{(1)} = \frac{\partial  C^{(1)}_1}{\partial r_1} +  C^{(1)}_1C^{(1)}_2 - C^{(1)}_3, \qquad k^{(1)} =  \frac{\partial  C^{(1)}_2}{\partial r_2} +  C^{(1)}_1C^{(1)}_2 - C^{(1)}_3, \\
h^{(2)} = \frac{\partial  C^{(2)}_1}{\partial R_1} +  C^{(2)}_1C^{(2)}_2 - C^{(2)}_3, \qquad k^{(2)} =  \frac{\partial  C^{(2)}_2}{\partial R_2} +  C^{(2)}_1C^{(2)}_2 - C^{(2)}_3
\end{gather*}
are related by the following equalities
\begin{gather*}
h^{(1)}(r_1,r_2) = \frac{d f_1}{d r_1} \frac{d f_2}{d r_2} h^{(2)}(f_1(r_1), f_2(r_2)), \qquad
k^{(1)}(r_1,r_2) = \frac{d f_1}{d r_1} \frac{d f_2}{d r_2} k^{(2)}(f_1(r_1), f_2(r_2)).
\end{gather*}

In particular, two equations are equivalent up to the factor-function  $w(r_1,r_2)$ only, if\/f
\[
h^{(1)}(r_1,r_2) = h^{(2)}(R_1,R_2), \qquad k^{(1)}(r_1,r_2) = k^{(2)}(R_1,R_2).
\]
Comparing the Laplace invariants of (\ref{x}) and (\ref{phi})
\begin{gather*}
h^{(x)} = -\frac{1}{\lambda_1 - \lambda_2} \frac{\partial^2 \lambda_2}{\partial r_1 \partial r_2} -\frac{1}{(\lambda_1 - \lambda_2)^2} \frac{\partial \lambda_2}{\partial r_1} \frac{\partial \lambda_2}{\partial r_2} = k^{(\varphi)}, \nonumber\\
k^{(x)} = \frac{1}{\lambda_1 - \lambda_2} \frac{\partial^2 \lambda_1}{\partial r_1 \partial r_2} -\frac{1}{(\lambda_1 - \lambda_2)^2} \frac{\partial \lambda_1}{\partial r_1} \frac{\partial \lambda_1}{\partial r_2} = h^{(\varphi)}, \nonumber
\end{gather*}
one can obtain the relation between eigenvalues
\begin{gather}
\label{lambda}
(\lambda_1 - \lambda_2) \frac{\partial^2 }{\partial r_1 \partial r_2} (\lambda_1 + \lambda_2) =  \frac{\partial \lambda_1}{\partial r_1} \frac{\partial \lambda_1}{\partial r_2} -  \frac{\partial \lambda_2}{\partial r_1} \frac{\partial \lambda_2}{\partial r_2}.
\end{gather}

Function $w$ can be determined from the following system of equations
\begin{gather}
\frac{1}{w}\frac{\partial w}{\partial r_1} = C_2^{(x)} - C_2^{(\varphi)}, \qquad \frac{1}{w}\frac{\partial w}{\partial r_2} = C_1^{(x)} - C_1^{(\varphi)}, \nonumber\\
\frac{\partial^2 w}{\partial r_1\partial r_2} + C_1^{(\varphi)} \frac{\partial w}{\partial r_1} + C_2^{(\varphi)} \frac{\partial w}{\partial r_2} = 0,\label{w}
\end{gather}
where $C_j^{(x)}$, $C_j^{(\varphi)}$ are the corresponding coef\/f\/icients in equations (\ref{x}) and (\ref{phi}).

Let us note, that a similar comparison of Laplace invariants one can make for the equations, obtained from (\ref{CL2}) and (\ref{hodograph}), but expressing functions $\psi$ and $y$.
\end{proof}

In particular, two simplest cases can be picked out:
\begin{enumerate}\itemsep=0pt
\item[1)] if $\det(\Lambda) = K = \text{const}$, then $\varphi = y(r_1,r_2)$, $\psi = K x(r_1,r_2)$;
\item[2)] if $\lambda_1 = -\lambda_2$, then $\varphi = x(r_1,r_2)$, $\psi = - y(r_1,r_2)$.
\end{enumerate}

An example of the f\/irst case is the system of the ideal plane plasticity \cite{Kachanov:2004}
\begin{gather*}
 M =  \begin{pmatrix}-\cot 2 v & -2k/\sin 2 v \\-1/(2k\sin2 v) & -\cot 2 v \end{pmatrix},
\end{gather*}
with $\det (\Lambda) = -1$.

One of the examples of system (\ref{eq00}) for the second case mentioned above, is system
\begin{gather*}
 m_{11} = m_{22} = 0, \qquad m_{12} = -1, \qquad m_{21} = F^2(u),
\end{gather*}
related to the well known equation
\begin{gather}
\label{Fermi}
\frac{\partial^2 Y}{\partial x^2} = F^2\left( \frac{\partial Y}{\partial y} \right) \frac{\partial^2 Y}{\partial y^2},
\end{gather}
investigated in~\cite{Zabusky:1962} in the context with Fermi, Pasta and Ulam~(1955) results on the vibration of a nonlinear, loaded (or beaded) f\/inite string. In such a case we have $\lambda_1 = -\lambda_2 = - F$.

We consider these and some other examples in the next section.

\section{Examples of application}
\label{sec:Ex}

It is well known, that linear homogeneous hyperbolic equations~(\ref{0}) are completely classif\/ied~\cite{Ovsyannikov} with respect to the group of admitted point symmetries. There are three dif\/ferent kinds of such equations, depending on the form of its Laplace invariants. As for Goursat problem~(\ref{adjoint}) for Riemann function, it can be reduced to the solution of an ordinary dif\/ferential equation in the same classif\/ied cases~\cite{Ibragimov}. Let us consider some examples of these kinds of equations.

\subsection[Plane ideal plasticity with Saint-Venant-Mises yield criterion]{Plane ideal plasticity with Saint-Venant--Mises yield criterion}
\label{subsec:1}
This system was investigated, using the group of admitted symmetries: for its invariant solutions see~\cite{ABS}, all its conservation laws and highest symmetries were described in~\cite{Senashov:1988} and for the reproduction of solutions by point transformations see~\cite{Senashov:2007, Yakhno:2009, Yakhno:2010}. Being semi-inverse method, group analysis provides analytical solutions and then one can determine the boundary conditions for obtained solutions. But if the the boundary conditions are given from the beginning, then the method of conservation laws can be applied to solve the boundary problem directly.

The system has the form
\begin{gather}
\frac{\partial \sigma}{\partial x}  - 2k\left( \frac{\partial\theta}{\partial x}  \cos 2\theta + \frac{\partial\theta}{\partial y} \sin 2\theta\right) = 0,\qquad
\frac{\partial \sigma}{\partial y}  - 2k\left( \frac{\partial \theta}{\partial x} \sin 2\theta - \frac{\partial\theta}{\partial y}  \cos
2\theta\right) = 0,\label{Senashov:eq1}
\end{gather}
where $\sigma$ is a hydrostatic pressure, $\theta$ is an angle
between the f\/irst main direction of a stress tensor and the
$Ox$-axis, $k$ is a constant of plasticity.

The functions for this system in the form (\ref{Riemann}) are as follows
\begin{gather*}
u=\sigma,\qquad v = \theta; \qquad \lambda_1 = \tan v,\qquad  \lambda_2 = - \cot v;\qquad  r_1 = \frac{u}{2k} - v,\qquad  r_2 = \frac{u}{2k} + v.
\end{gather*}

In this case $\det(\Lambda) = -1$, so $\varphi = y(r_1,r_2)$, $\psi = -x(r_1,r_2)$.

The solution of problem (\ref{CL2}), (\ref{init_conds_linear}) has a form \cite{Senashov:2004}
\begin{gather*}
\varphi = 2\frac{\partial \rho}{\partial r_1} \cos v -\rho \sin v, \qquad  \psi = 2\frac{\partial \rho}{\partial r_1} \sin v + \rho \cos v,
\end{gather*}
where function $\rho(r_1,r_2)$ looks like this
\[
\rho (r_1, r_2) = R\left(r_1,r_1^0,r_2,r_2^0\right)
 \cos \left (\frac {r_2^0- r_1^0} {2} \right)-\frac {1} {2} \int_{r_2^0}^{r_2}
  R\left(r_1,r_1^0,r_2,\tau\right) \sin\left (\frac {\tau-r_1^0}{2} \right) \, d\tau.
\]
Accordingly, the solution of the problem (\ref{CL2}),
(\ref{init_conds_linear_y}) is
\[
\rho (r_1, r_2) = R\left(r_1,r_1^0,r_2,r_2^0\right)
 \sin \left (\frac {r_2^0- r_1^0} {2} \right) + \frac {1} {2} \int_{r_2^0}^{r_2}
  R\left(r_1,r_1^0,r_2,\tau\right) \cos\left (\frac {\tau-r_1^0}{2} \right) \, d\tau,
\]
$R\left(r_1,r_1^0,r_2,r_2^0\right)=I_0\left(\sqrt {(r_1-r_1^0) (r_2-r_2^0)}\right)$  is the modif\/ied Bessel function of the f\/irst kind of a zero order, having the following properties
\begin{gather*}
 I_0 (0) = 1,\qquad  I_0^\prime (0) = 0, \qquad \frac{d I_0(z)}{d z} = I_1(z), \qquad \frac{d I_1(z)}{d z} = I_0(z) - \frac{I_1(z)}{z}.
\end{gather*}

Let us solve one practical problem to show the ef\/fectivity of using conservation laws instead of linearization by hodograph transformation.

In work \cite{Khristianovich:1936} the Cauchy problem for the cavity in an inf\/inite medium loaded by a constant shear stress in addition to a uniform pressure (the so called Mikhlin problem)  was solved under the condition $J_1 \ne 0$. Using conservation laws permits to forget about this condition.

Let us consider an example, namely, let the contour be given by the $2\pi$-periodic curve
\[
	L= \begin{cases}
	x = - r \cot t,\qquad y = - r, \quad & t \in[ \gamma - \pi , - \gamma ),\\
	x = - r \cot t,\qquad  y = r,  \quad &  t \in (\gamma ,\pi - \gamma ), \\
  x = r \cos \dfrac{\pi t}{2\gamma},\qquad y = r\sin \dfrac{\pi t}{2\gamma}, \quad &  t \in [ - \gamma ,\gamma ], \\	
  x = r\cos \dfrac{\pi t}{2(\pi - \gamma )} - a,\qquad y = r\sin \dfrac{\pi t}{2(\pi - \gamma )}, \quad & t \in (\gamma - \pi ,\pi - \gamma ],
	\end{cases}
\]
where $a$ is the distance from $(0,0)$ to the center of curve parts with radio $r$, $\gamma = \arctan (r / a)$. This contour is loaded by the normal and tangent stresses
 \[
\sigma _n = - p, \qquad \tau _n = 0,
\]
where $n$ is the normal to the contour. Let us put $p = k = 1/2$ and def\/ine the angle between the tangent to the contour and $x$-axes as  $N(t) = \arctan \frac{y'_t} {x'_t}$, then the initial conditions take the form
\[
\left. \sigma \right|_L = - 1, \qquad  \left. \theta \right|_L =
\begin{cases}
	N(t) - \pi / 4 + \pi / 2,\quad & t \in (0,\pi ),\\
	N(t) - \pi / 4 + 3\pi / 2, \quad & t \in (\pi ,2\pi ),\\
	- \pi / 4,\quad & t = 0,\\
	3\pi / 4,\quad & t = \pi.
\end{cases}
\]

\begin{figure}[t]
\centering
\includegraphics[scale=0.55]{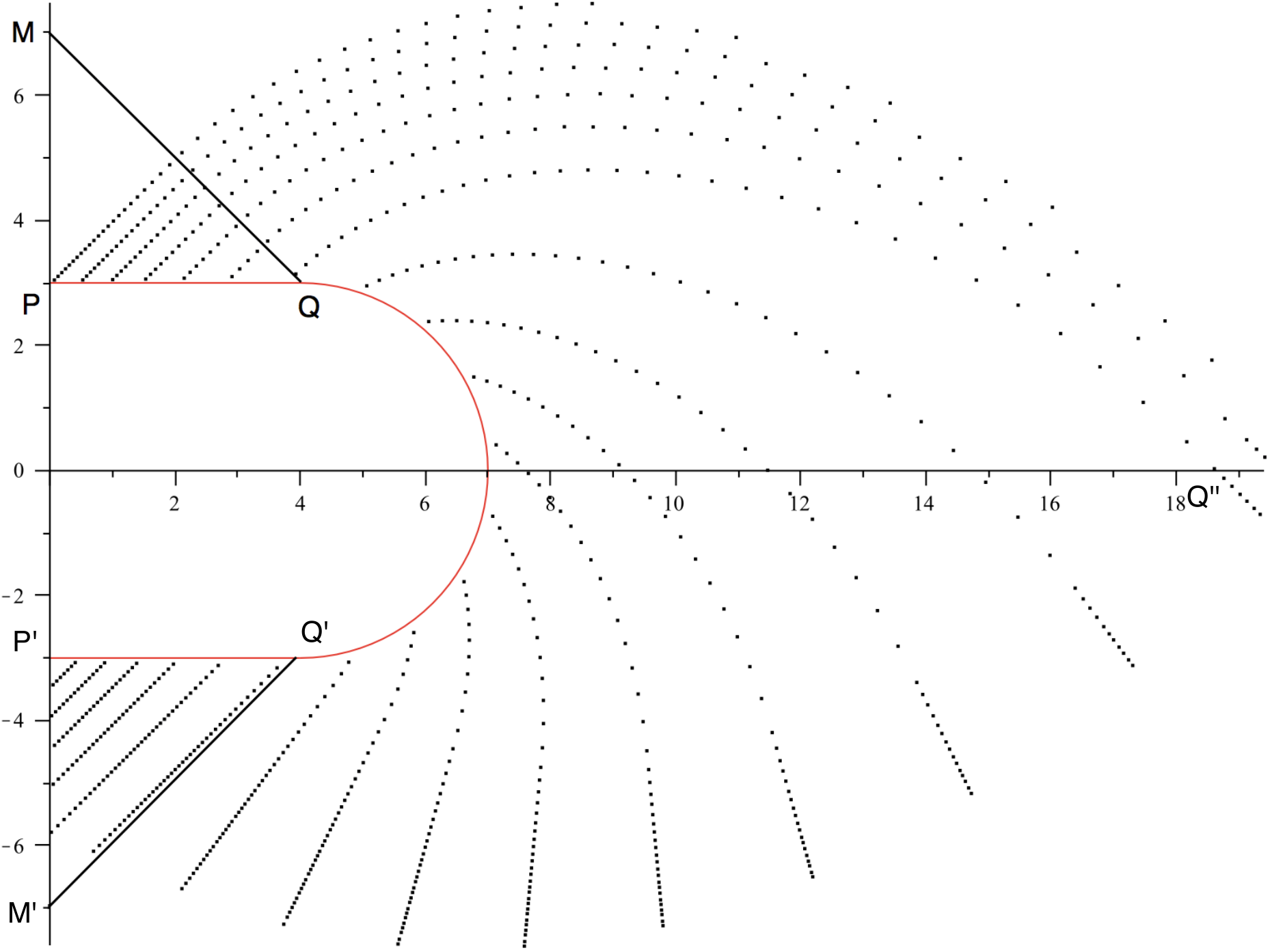}

\caption{Characteristic f\/ield for the cavity.}
\label{f3}
\end{figure}

The solution of this problem is given in Fig.~\ref{f3}. Here $a = 4$, $r = 3$ and the f\/irst family of the characteristic curves is constructed, using conservation laws described above.  Due to the symmetry of the contour, only the part with $x>0$ is shown.

Let us note, that  the stress f\/ield near the straight-line border of the cavity (regions $PQM$ and $P' Q' M'$) can not be constructed as a solution of a linearized system because  $J_1 = 0$ in this domain. Moreover, the ordinary way to solve this problem is to consider the dif\/ferent parts separately. Firstly, one has to construct the f\/ield in the region $PQM$, assuming $J_1=0$. Then one need to solve the Cauchy problem for the $QQ'$ segment of the cavity. Having the solution of above problems, one should construct the f\/ield in the $MQQ''$ region as a solution of the characteristic boundary problem, when the initial data are given along two characteristic curves~$MQ$ and~$QQ''$. Using the conservation laws permits to avoid such a division of the boundary line and to solve the problem of Cauchy for the whole contour.

\subsection{Plane plasticity with Coulomb yield criterion}

The generalization of (\ref{Senashov:eq1}) is a system of the plane isotropic soil plasticity equations under Coulomb yield criterion, which in the case when the weight of the soil is neglected, has the form~\cite{Chakrabarty:2006, Sokolovskii:1960}
\begin{gather*}
 \frac{\partial \sigma}{\partial x} (1+\cos 2\alpha \cos 2\Theta)  +  \frac{\partial \sigma}{\partial y} \cos 2\alpha \sin 2\Theta   \\
\qquad{} =  2(\sigma \cos 2\alpha + k\sin 2\alpha)\left(\frac{\partial \Theta}{\partial x} \sin 2\Theta -
\frac{\partial \Theta}{\partial y} \cos 2\Theta\right),    \\
\frac{\partial \sigma}{\partial x} \cos 2\alpha \sin 2\Theta  +  \frac{\partial \sigma}{\partial y} (1 - \cos 2\alpha
\cos 2\Theta)   \\
  \qquad{} =  - 2(\sigma \cos 2\alpha + k\sin 2\alpha)\left(\frac{\partial \Theta}{\partial x}
\cos 2\Theta +\frac{\partial \Theta}{\partial y} \sin 2\Theta\right),
\end{gather*}
where $\pi/2 - 2\alpha$ is a constant angle of internal friction, $\alpha\in\left(0,\frac{\pi}{2}\right)$,
$\Theta = \theta+ \pi/4$ and $k$ denotes the cohesion of the soil. If $\alpha = \pi/4$, then we have system~(\ref{Senashov:eq1}).

The Riemann invariants and the corresponding eigenvalues are the following ones $(\alpha \ne \pi/4)$
\begin{gather*}
\sigma = u,\qquad \theta = v, \qquad \lambda_{1,2} =  \tan(\Theta  \pm \alpha), \qquad
r_{1,2} = \frac{\tan 2\alpha}{2}  \ln(\sigma\cot 2\alpha +k) \pm \Theta.
\end{gather*}

If we take
\begin{gather*}
R_1 =  -r_1, \qquad R_2 = - r_2, \qquad \varphi(r_1, r_2) = y(R_1, R_2), \qquad \psi(r_1,r_2) =  -x(R_1,R_2),
\end{gather*}
then equations for conservation laws (\ref{CL2})
\begin{gather}
\label{Coulomb}
 \frac{\partial \psi}{\partial r_1}   - \tan\left(\frac{r_1-r_2}{2} + \alpha\right) \frac{\partial   \varphi}{\partial r_1} = 0, \qquad
 \frac{\partial \psi}{\partial r_2}   - \tan\left(\frac{r_1-r_2}{2} - \alpha\right) \frac{\partial \varphi}{\partial r_2} = 0
\end{gather}
coincide with linearized ones (\ref{hodograph})
\begin{gather*}
 \frac{\partial y}{\partial R_1}   - \tan\left(\frac{R_1-R_2}{2} - \alpha\right) \frac{\partial  x}{\partial R_1} = 0, \qquad
 \frac{\partial y}{\partial R_2}   - \tan\left(\frac{R_1-R_2}{2} + \alpha\right) \frac{\partial  x}{\partial R_2} = 0.
\end{gather*}
This is a simple example, when one can need a change of independent variables $r_1$, $r_2$ to relate the conservation laws to the solution of a linearized system.

Let us note, that introducing new functions $\Phi(r_1,r_2)$ and $\Psi(r_1,r_2)$ in such a way
\begin{gather*}
\varphi  =  \frac{ e^{\cot 2\alpha (r_1+r_2)/2}}{\sin 2\alpha}\left(-\Phi \cos(\Theta + \alpha) + \Psi \cos(\Theta - \alpha) \right), \nonumber \\
\psi  =  \frac{ e^{\cot 2\alpha (r_1+r_2)/2}}{\sin 2\alpha}\left(-\Phi \sin(\Theta + \alpha) + \Psi \sin(\Theta - \alpha) \right)
\end{gather*}
one can reduce system (\ref{Coulomb}) to the following one
\begin{gather*}
\frac{\partial \Phi}{\partial r_2} + \frac{\Psi}{2\sin 2\alpha} = 0, \qquad \frac{\partial \Psi}{\partial r_1} + \frac{\Phi}{2\sin 2\alpha} = 0
\end{gather*}
and f\/inally to the telegraph equation of the form
\begin{gather}
\label{tele}
\frac{\partial^2\Psi}{\partial r_1\partial r_2} - \frac{\Psi}{4\sin^2 2\alpha} = 0.
\end{gather}

The corresponding initial conditions (\ref{init_conds_linear}) look like
\begin{gather}
\label{tele_x}
\left.\Psi\right|_{r_1 = r_1^0} = - e^{-\gamma(r_1^0 +r_2)} \cos\left(\frac{r_1^0-r_2}{2} + \alpha \right),\qquad  \left.\frac{\partial \Psi}{\partial r_1}\right|_{r_2 = r_2^0} =0,
\end{gather}
and for (\ref{init_conds_linear_y}) we have
\begin{gather}
\label{tele_y}
\left.\Psi\right|_{r_1 = r_1^0} = - e^{-\gamma(r_1^0 +r_2)} \sin\left(\frac{r_1^0-r_2}{2} + \alpha \right),\qquad  \left.\frac{\partial \Psi}{\partial r_1}\right|_{r_2 = r_2^0} =0,
\end{gather}
where $\gamma = \frac{1}{2}\cot 2\alpha$.

The solution of problem (\ref{tele}), (\ref{tele_x}) has a form
\begin{gather*}
\Psi  =  - e^{-\gamma (r_1^0+r_2^0)} \cos\left(\frac{r_1^0-r_2^0}{2} + \alpha \right) R\big(r_1,r_1^0,r_2,r_2^0\big)     \\
\hphantom{\Psi  =}{} +  \int_{r_2^0}^{r_2} R\big(r_1,r_1^0,r_2,\tau\big) e^{-\gamma(r_1^0+\tau)} \left[ \gamma \cos\left ( \frac{r_1^0-\tau}{2}+\alpha\right) - \frac{1}{2} \sin\left( \frac{r_1^0 - \tau}{2}+\alpha \right) \right] \, d\tau,
\end{gather*}
accordingly, the solution for (\ref{tele}), (\ref{tele_y}) is as follows
\begin{gather*}
\Psi  =  - e^{-\gamma (r_1^0+r_2^0)} \sin\left(\frac{r_1^0-r_2^0}{2} + \alpha \right) R\big(r_1,r_1^0,r_2,r_2^0\big)     \\
\hphantom{\Psi  =}{}  +  \int_{r_2^0}^{r_2} R\big(r_1,r_1^0,r_2,\tau\big) e^{-\gamma(r_1^0+\tau)} \left[ \gamma \sin \left ( \frac{r_1^0-\tau}{2}+\alpha\right) + \frac{1}{2} \cos\left( \frac{r_1^0 - \tau}{2}+\alpha \right) \right] \, d\tau,
\end{gather*}
where $R\left(r_1,r_1^0,r_2,r_2^0\right)=I_0\left(\frac{1}{\sin 2\alpha}\sqrt {(r_1-r_1^0) (r_2-r_2^0)}\right)$.

\subsection{Nonlinear hyperbolic heat equation}
As indicated in \cite{Curro:2008}, the hyperbolic heat equation
\begin{gather*}
\frac{\partial}{\partial t}\left(\frac{\partial U}{\partial t} + \frac{U}{\tau_0}\right) - \frac{\partial}{\partial X}\left( \frac{\chi_0^2}{U^2} \frac{\partial U}{\partial X}\right) = 0,
\end{gather*}
where $\chi_0$ and $\tau_0$ are positive constants can be expressed in the form of the following quasilinear system
\begin{gather*}
\frac{\partial u}{\partial x} - \frac{\partial v}{\partial y}=0, \qquad
\frac{\partial v}{\partial x} - \frac{\chi_0^2}{u^2} \frac{\partial u}{\partial y}=0
\end{gather*}
by introducing the potential function $v(t,X)$ and setting
\[
U=\tau_0 e^{-t/\tau_0}u, \qquad y=X, \qquad x=e^{t/\tau_0}.
\]
In this case
\[
r_{1,2} = u e^{ \mp v/\chi_0 },\qquad  \lambda_1 = - \lambda_2 = \frac{\chi_0}{\sqrt{r_1r_2}},\qquad u>0,
\]
so $\varphi = x(r_1,r_2)$, $\psi = - y(r_1,r_2)$.

Let us note, that in this case, equation (\ref{phi}) can be reduced to the following one
\begin{gather*}
\frac{\partial^2 \varphi}{\partial x_1 \partial x_2}  -  \frac{\partial \varphi}{\partial x_1} - \frac{\partial \varphi}{\partial x_2}=0,
\end{gather*}
introducing new variables $r_i = e^{4 x_i}$.

Explicitly, solution of problem  (\ref{CL2}), (\ref{init_conds_linear}) looks like this
\begin{gather*}
\varphi  =  \frac{2}{\chi_0} r_1^{\frac{3}{2}} r_2^{\frac{1}{2}} \frac{\partial \Phi}{\partial r_1},\qquad  \psi = 2 r_1\frac{\partial \Phi}{\partial r_1} + \Phi, \\
\Phi  =  \left(\frac{r_1^0}{r_1 r_2}\right)^{\frac{1}{4}} \left[ \big(r_2^0\big)^{\frac{1}{4}} I_0\left(\frac{1}{2} \ln^{\frac{1}{2}}\frac{r_1}{r_1^0} \ln^{\frac{1}{2}}\frac{r_2}{r_2^0}\right) + \frac{1}{4} \int_{r_2^0}^{r_2}   I_0\left(\frac{1}{2} \ln^{\frac{1}{2}}\frac{r_1}{r_1^0} \ln^{\frac{1}{2}}\frac{r_2}{t}\right) t^{-\frac{3}{4}} dt\right].
\end{gather*}
For problem (\ref{CL2}), (\ref{init_conds_linear_y}) function $\Phi$ is the following one
\begin{gather*}
\Phi = \frac{\chi_0}{\left({r_1^0}{r_1 r_2}\right)^{\frac{1}{4}}}\left[ \big(r_2^0\big)^{-\frac{1}{4}} I_0\left(\frac{1}{2} \ln^{\frac{1}{2}}\frac{r_1}{r_1^0} \ln^{\frac{1}{2}}\frac{r_2}{r_2^0}\right) - \frac{1}{4}  \int_{r_2^0}^{r_2}   I_0\left(\frac{1}{2} \ln^{\frac{1}{2}}\frac{r_1}{r_1^0} \ln^{\frac{1}{2}}\frac{r_2}{t}\right) t^{-\frac{5}{4}} dt \right] .
\end{gather*}

\subsection{Gas dynamics}
 The one-dimensional isentropic f\/low of polytropic gas in Euler coordinates in case of plane symmetries, as is well known~\cite{Rozhdestvenskii:1983}, is described by the following hyperbolic system
\begin{gather}
\label{gas}
\frac{\partial s}{\partial t} + (\alpha s + \beta r) \frac{\partial s}{\partial x} = 0, \qquad
\frac{\partial r}{\partial t} + (\alpha r + \beta s) \frac{\partial r}{\partial x} = 0,
\end{gather}
where $s = r_1$, $r = r_2$ are the Riemann invariants, so
\[
\lambda_1 = \alpha r_1 + \beta r_2,\qquad \lambda_2 = \alpha r_2 + \beta r_1,
\]
$\alpha = 1/2 + (\gamma - 1)/4$, $\beta=1/2 - (\gamma - 1)/4$, $\gamma = \text{const} \ne \pm 1$ is a parameter of polytrope.

Explicitly, the solution of the problem (\ref{CL2}), (\ref{init_conds_linear}) has the form
\begin{gather*}
\psi= \frac{\lambda_1 \rho_1 - \lambda_2 \rho_2}{\lambda_1 - \lambda_2}, \qquad  \varphi = \frac{\rho_1 - \rho_2}{\lambda_1 - \lambda_2},
\end{gather*}
where
\begin{gather*}
\rho_1 = \rho_2 - \frac{r_2 - r_1}{K} \frac{\partial \rho_2}{\partial r_1}, \qquad K=\frac{\gamma + 1}{2(1-\gamma)},
\\
\rho_2(r_1,r_2)  =  \frac{(r_2^0 - r_1^0)^{2K + 1}}{(r_2^0 - r_1)^{K} (r_2 - r_1^0)^{(K+1)}}  F\big(r_1^0,r_2^0;r_1,r_2\big)  \\
\hphantom{\rho_2(r_1,r_2)  =}{} + \frac{K+1}{(r_2 - r_1^0)^{(K+1)}} \int_{r_2^0}^{r_2} \frac{(t-r_1^0)^{2K}}{(t-r_1)^{K} } F\big(r_1^0,t; r_1,r_2\big) dt,
\end{gather*}
and $F(r_1^0,r_2^0;r_1,r_2) =  {}_2F_1\left( K, K+1; 1; \frac{(r_1^0 - r_1)(r_2 - r_2^0)}{(r_2 - r_1^0)(r_2^0 - r_1)}\right)$ is a hypergeometric function.

Analogically, for problem (\ref{CL2}), (\ref{init_conds_linear_y}) function $\rho_2$ undergoes a little modif\/ication
\begin{gather*}
 \rho_2(r_1,r_2) = \frac{(\beta r_2^0 + \alpha r_1^0)(r_2^0 - r_1^0)^{2K + 1}}{(r_2^0 - r_1)^{K} (r_2 - r_1^0)^{(K+1)}}
  F\big(r_1^0,r_2^0;r_1,r_2\big)  \\
\hphantom{\rho_2(r_1,r_2) =}{} +  \frac{1}{(r_2 - r_1^0)^{(K+1)}} \int_{r_2^0}^{r_2} \frac{(t-r_1^0)^{2K+1}}{(t-r_1)^{K} } F\big(r_1^0,t; r_1,r_2\big) \left(\beta + \frac{K+1}{t - r_1^0}(\beta t + \alpha r_1^0)\right)dt.
\end{gather*}

In this case, equation (\ref{lambda}) is satisf\/ied, because $h^{(x)} = h^{(\varphi)} = -\alpha \beta$, so there is a relation between the conservation laws and solutions of linearized system of the form $\varphi = w(r_1,r_2) x(r_1,r_2)$, where function~$w$ is as follows
\begin{gather*}
w = (r_1 - r_2)^{(\alpha + \beta)/(\alpha - \beta)} = (r_1 - r_2)^{2/(\gamma-1)}.
\end{gather*}

\subsection{Loaded homogeneous semi-inf\/inite elastic-plastic beam}

The process of propagation of plastic deformations in semi-inf\/inite elastic-plastic beam, dy\-na\-mi\-cally loaded on one end, in Lagrange coordinates is described by the following system \cite{Nowacki:1978}
\begin{gather*}
\rho \frac{\partial v}{\partial t} = \frac{\partial \sigma}{\partial x},\qquad  \frac{\partial v}{\partial x} = \frac{1}{\rho a^2(\sigma)} \frac{\partial \sigma}{\partial t},
\end{gather*}
where $\rho = \text{const}$ is a density, $v(t,x)$ is a velocity of medium particles, a tension $\sigma = \sigma(\varepsilon)$ is monotonically increasing convex function of the deformation $\varepsilon(x,t)$. Introducing function $u(x,t)$, so that $v=\frac{\partial u}{\partial t}$, $\varepsilon = \frac{\partial u}{\partial x}$, we come to nonlinear wave equation
\[
\frac{\partial^2 u}{\partial t^2} - a^2(\varepsilon) \frac{\partial^2 u}{\partial x^2} =0,
\]
so $a(\varepsilon)$ is a speed of the longitudinal wave propagation in the beam. Note, that the above equation is of type~(\ref{Fermi}).

Eigenfunctions have the form $ \lambda_{1,2} = \mp a(\sigma)$ and Riemann invariants are the following ones
\begin{gather*}
r_{1,2} = v \pm \int_0^\sigma \frac{d\sigma_1}{a(\sigma_1)}.
\end{gather*}
This is the case when $\lambda_1 = -\lambda_2$, then $\varphi = t(r_1,r_2)$, $\psi = - x(r_1,r_2)$.

For example, if $a = \sqrt{\sigma}$, then
\[
\lambda_{1,2} = \mp \sqrt{\sigma}, \qquad r_{1,2} = v \pm 2\sqrt{\sigma},
\]
and system (\ref{CL2}) can be reduced to Euler--Poisson--Darboux equation in the form
\begin{gather*}
 \frac{\partial^2 \varphi}{ \partial r_1 \partial r_2} + \frac{2}{r_1 - r_2} \left( -\frac{\partial \varphi}{\partial r_1} +\frac{\partial \varphi}{\partial r_2}  \right) = 0
\end{gather*}
with the well known Riemann function.

Moreover, this case is reduced to the system \eqref{gas} with
\begin{gather*}
s = -r_1,\qquad r = - r_2,\qquad K=-1/2,\qquad \alpha = -1/4,\qquad \beta = 1/4.
\end{gather*}

Solution of the problem (\ref{CL2}), (\ref{init_conds_linear}) has the form
\begin{gather*}
\varphi = 2\frac{\rho_1 - \rho_2}{r_1-r_2}, \qquad \psi = \frac{\rho_1 + \rho_2}{2},\qquad \rho_1 = \rho_2 - 2(r_1-r_2) \frac{\partial \rho_2}{\partial r_1},
\\
\rho_2(r_1,r_2) = \left( \frac{r_1 - r_2^0}{r_1^0 - r_2}\right)^{\frac{1}{2}}   F\big(r_1^0,r_2^0;r_1,r_2\big)
- \frac{1}{2\sqrt{r_1^0 - r_2}} \int_{r_2^0}^{r_2} \frac{\sqrt{r_1 - t}}{r_1^0 - t} F\big(r_1^0,t; r_1,r_2\big) dt,
\end{gather*}
where $F(r_1^0,r_2^0;r_1,r_2) = {}_2F_1\left( -1/2, 1/2; 1; \frac{(r_1-r_1^0)(r_2 - r_2^0)}{(r_1 - r_2^0)(r_2 - r_1^0)}\right)$ is a hypergeometric function.

Function $\rho_2$  for problem (\ref{CL2}), (\ref{init_conds_linear_y}) is the following one
\begin{gather*}
\rho_2(r_1,r_2) = \frac{r_1^0 - r_2^0}{4} \left( \frac{r_1 - r_2^0}{r_1^0 - r_2}\right)^{\frac{1}{2}}   F\big(r_1^0,r_2^0;r_1,r_2\big)\\
 \hphantom{\rho_2(r_1,r_2) =}{}
 - \frac{3}{8\sqrt{r_1^0 - r_2}} \int_{r_2^0}^{r_2} \sqrt{r_1 - t}   F\big(r_1^0,t; r_1,r_2\big) dt.
\end{gather*}

\subsection[The Born-Infeld equation]{The Born--Infeld equation}

Let us consider one of the representations of  Born--Infeld \cite{Born:1934} nonlinear electrodynamics model in the form of linearly polarized plane wave equation
\begin{gather*}
\left[ 1+ \left(\frac{\partial w}{\partial x}\right)^2\right] \frac{\partial^2 w}{\partial t^2} - 2 \frac{\partial w}{\partial x} \frac{\partial w}{\partial t} \frac{\partial^2 w}{\partial x \partial t} - \left[ 1- \left(\frac{\partial w}{\partial t}\right)^2\right] \frac{\partial^2 w}{\partial x^2} =0,
\end{gather*}
which is a hyperbolic one for the solutions with $1 + \left(\frac{\partial w}{\partial x}\right)^2 - \left(\frac{\partial w}{\partial t}\right)^2>0$. Introducing
\begin{gather*}
u = \frac{\partial w}{\partial x},\qquad  v = \frac{\partial w}{\partial t}
\end{gather*}
one can obtain the system in the form (\ref{eq00})
\begin{gather*}
\frac{\partial u}{\partial t} - \frac{\partial v}{\partial x} = 0,\qquad
\frac{\partial v}{\partial t} + \frac{v^2-1}{1+u^2} \frac{\partial u}{\partial x} - \frac{2uv}{1+u^2} \frac{\partial v}{\partial x} =0.
\end{gather*}
The Riemann invariants and eigenvalues are the following ones
\begin{gather*}
r_1 =\frac{-uv - \sqrt{1+u^2-v^2}}{1+u^2}, \qquad r_2 = \frac{-uv + \sqrt{1+u^2-v^2}}{1+u^2}, \qquad \lambda_1 = r_2, \qquad \lambda_2 = r_1,
\end{gather*}
and the corresponding system has the form  ($r_1 = r$, $r_2 = s$)
\begin{gather*}
\frac{\partial r}{\partial t} + s \frac{\partial r}{\partial x} = 0, \qquad
\frac{\partial s}{\partial t} + r \frac{\partial s}{\partial x} = 0.
\end{gather*}
The above system is a special case of (\ref{gas}) with $\gamma = -1$, $\alpha = 0$, $\beta = 1$, i.e.\ the so-called Chaplygin gas.

Note, that this case is an exceptional one, because Laplace invariants are zeros both for (\ref{x}) and for (\ref{phi}),
which take the form
\begin{gather*}
\frac{\partial^2 x}{\partial r_1 \partial r_2} = 0, \qquad  \frac{\partial^2 \varphi}{\partial r_1 \partial r_2} - \frac{1}{r_1 - r_2} \frac{\partial \varphi}{\partial r_1} + \frac{1}{r_1 - r_2} \frac{\partial \varphi}{\partial r_2} = 0.
\end{gather*}
Solving (\ref{w}), we obtained the relation of the components of conservation law $\varphi$, $\psi$ with the solution of linearized system
\begin{gather*}
\varphi(r_1,r_2) = -\frac{ x(r_1,r_2)}{r_1 - r_2}, \qquad \psi(r_1,r_2) = \frac{r_1 r_2}{r_1 - r_2} y(r_1,r_2).
\end{gather*}
Let us note, that the above conservation laws are the particular case of ones calculated in~\cite{Menshikh:1989}.

Explicitly, the solution of problem (\ref{CL2}), (\ref{init_conds_linear}) has a form
\begin{gather}
\label{BI_x}
\varphi = \frac{1}{r-s},\qquad \psi = \frac{r}{r-s},
\end{gather}
and for the problem (\ref{CL2}), (\ref{init_conds_linear_y}) we have
\begin{gather}
\label{BI_y}
\varphi = \frac{s}{r-s},\qquad \psi = \frac{sr}{r-s}.
\end{gather}

The f\/inal remark is that solution~(\ref{sol_x}), (\ref{sol_y}), corresponding to~(\ref{BI_x}) and~(\ref{BI_y}) accordingly,  is equivalent to the one provided in~\cite{Barbishov:1966} under the condition for Jacobian $J_1 \ne 0$,  and expressed in~\cite{Menshikh:1989} in terms of Riemann invariants for the initial value problem: $r|_{t=0} = r^0(x)$, $s|_{t=0} = s^0(x)$.

\section{Conclusions}\label{sec:C}

Application of Riemann method to the system, describing conservation laws of quasilinear system, instead of linearized system, permits to solve the boundary value problems in the domain, where Jacobian of hodograph transformation is equal to zero. The  proposed method permits the construction of characteristics, including both straight lines, corresponding to the simple wave solution (the constant solution too), and the case, when the initial curve is an envelope of a family of characteristic curves, which is useful in applications.

In our paper is shown how the conservation laws (of the special form) of the considered system could be applied to construct the solution of a Cauchy problem, without separation of regions with a simple wave  (or constant) solution, i.e.\  permit to f\/ind the point of characteristic without any restriction: is this point of straight line or not. As far we know it is a new point in the theory of considered form of quasilinear systems.

The obtained theorem relates the Cauchy problem for {quasilinear} system with the solution of Cauchy problem for {linear} system of conservation laws.

The relation between conservation laws of a quasilinear system and the corresponding li\-nea\-ri\-zed system, obtained by the hodograph transformation, gives nonsingular solutions.

Recently, in~\cite{Curro:2008} some non-homogeneous quasilinear systems, arising in various areas of physical interest, were reduced to the homogeneous ones by using the invariance to suitable Lie groups of point transformations. The well-known relation between the symmetries and conservation laws, we hope, opens the way to apply the proposed method for some non-homogeneous systems.

\subsection*{Acknowledgments}
We would like to express our gratitude to unknown referees for useful corrections. This work was partially supported by PRO-SNI (UdeG).

\pdfbookmark[1]{References}{ref}
\LastPageEnding

\end{document}